# Mechanical and Vibrational Characteristics of Functionally Graded Cu-Ni Nanowire: A Molecular Dynamics Study

Mahmudul Islam[1], Md Shajedul Hoque Thakur[1], Satyajit Mojumder[1,2], Abdullah Al Amin[3] and Md Mahbubul Islam[4*]

[1]Department of Mechanical Engineering, Bangladesh University of Engineering & Technology (BUET), Dhaka, 1000, Bangladesh

[2]Theoretical and Applied Mechanics Program, Northwestern University, Evanston, IL-60208, USA

[3]Department of Mechanical and Aerospace Engineering, Case Western Reserve University, 10900 Euclid Avenue, Cleveland, OH 44106-7222, USA

[4] Department of Mechanical Engineering, Wayne State University, 5050 Anthony Wayne Drive, Detroit, MI 48202, USA

[*]Corresponding author, *Email address:* mahbub.islam@wayne.edu

# Abstract

Functionally graded material (FGM) is a class of advanced materials, consisting of two (or more) different constituents, that possesses a continuously varying composition profile. With the advancement of nanotechnology, applications of FGMs have shifted from their conventional usage towards sophisticated micro and nanoscale electronics and energy conversion devices. Therefore, the study of mechanical and vibrational properties of different FGM nanostructures is crucial in exploring their feasibility for different applications. In this study, for the first time, we employed molecular dynamics (MD) simulations to investigate the mechanical and vibrational properties of radially graded *Cu-Ni* FGM nanowires (NW). Distribution of *Cu* and *Ni* along the radial direction follows power-law, exponential and sigmoid functions for FGM NWs under consideration. Our results demonstrate that, distribution function parameters play an important role in modulating the mechanical (elastic modulus and ultimate tensile strength) and vibrational (natural frequency and quality factor) properties of FGM NWs. The study also suggests that, elastic moduli of FGM NWs can be predicted with relatively good accuracy using Tamura and Reuss micromechanical models, regardless of NW diameter. We found that, Euler-Bernoulli beam theory under-predicts the natural frequencies of FGM NWs, whereas He-Lilley model closely approximates the MD results. Interestingly, FGM NWs are always found to exhibit beat vibration because of their asymmetrical cross sections. Finally, this is the first atomistic scale study of FGMs that directly compares MD simulations with continuum theories and micromechanical models to understand the underlying mechanisms that govern the mechanical and vibrational properties of FGM NWs in nanoscale.

# 1. Introduction

Functionally graded materials (FGM) are considered one of the most promising materials among the class of advanced fabricated materials. They consist of two or more different materials, where the composition continuously varies along a dimension following a particular function.[1] FGMs are conceived as a solution to solve high-stress concentration, high-temperature creep and material delamination[2] challenges that are common in other fabricated materials such as composites. These enhanced thermal and mechanical properties render FGM a suitable candidate for manufacturing structures of aeroplane, automobile engine components and protective coatings for turbine blades.[3,4] Furthermore, recent researches showed that, human teeth and bone can be considered as functionally graded living tissues from nature.[5] Thus, application of FGMs have found its way into teeth and bone replacement industry. Due to the applications in the fields of aerospace, automobile, medicine and energy, the research efforts to characterize the mechanical and thermal properties of FGMs have increased rapidly in recent years.

Numerous research works have been conducted to understand the mechanical and vibrational properties of FGMs with the help of well-established theories of continuum mechanics and finite element method (FEM).[6–12] Karman theory has been applied for large deformation in order to get an analytical solution for FGM plates under transverse mechanical loading.[10] Recently an efficient higher order shear and normal deformation theory of FGM plates has been presented where the number of unknown functions is only five as opposed to six or more in previous studies.[11] A formulation of closed form analytical expression of FGM cylinders has also been established using Variational Asymptotic Method (VAM).[12] Finite element method (FEM) in conjunction with continuum mechanics is also extensively utilized to understand various mechanical and vibrational properties of FGMs. The nonlinear thermo-elasticity of functionally graded ceramic-metal plates has been described using plate finite element method that accounts for the transverse shear strains along with rotary inertia and moderately large rotations in the Von Karman sense.[13] He *et al.* used a finite element formulation, based on classical laminated plate theory to study the vibration control of FGM plates.[14] Researchers also investigated mechanical behavior of FGM nano-cylinders under radial pressure by FEM. These investigations demonstrated that, material inhomogeneity index has significant influence on radial and circumferential stresses.[15] The material properties and slenderness ratio of simply supported FGM beams are also found to play a significant role on their frequencies of free vibration.[16] These studies on the characteristics of FGMs are all based on either theoretical model or FEM which are related to classical mechanics approach.

In the recent years, applications of FGMs have shifted towards sophisticated micro and nanoscale electronics and energy conversion devices such as broadband ultrasonic transducer, solid oxide fuel cells,[17] high current connectors[18] and thermoelectric energy converter.[3] These uses of FGM nanostructures demand an extensive study of their mechanical, thermal and electronic properties in micro and nanoscale. Molecular dynamics (MD) simulations can be a very effective method to investigate and understand the mechanical and vibrational characteristics of FGMs at nanoscale. Several MD studies have been conducted to understand the mechanical and vibrational behaviors of nanowires (NWs), nanoplates and nanobeams made of different homogenous and non-homogenous composite materials. Koh *et al.* reported that, the elastic modulus of the nanowire is lower than its bulk counterpart.[19] This phenomenon have been explained through bond saturation. A surface of a nanowire can be either softer or stiffer than its bulk counterpart depending on the competition between atomic co-ordination and electronic redistribution. In the studies on NWs made of alloys, it is observed that the mechanical properties of NWs improve to a certain extent with increase of certain constituent fractions.[20] Additionally, various fundamental investigations on the vibrational properties of NWs have also been conducted. It is found that at high initial excitation the NW exhibits beat vibration.[21] Also the Elastic modulus obtained from MD computations, which is directly related to the natural frequency, overestimates the classical beam theory in case of alloy NWs.[22] These vibrational studies also showed that quality factor of the nanowire decreases with the increasing temperature.[22] However, to the best of authors' knowledge, no atomic scale study has been conducted to investigate the mechanical and vibrational properties of FGM nanostructures using MD simulation. This is due to the absence of a proper approach for modeling FGMs using MD simulations.

Along with the mechanical and vibrational properties, understanding of the mechanics of FGMs and optimum distribution profiles of the material constituent are also of great interest. To describe the variation of constituents of FGMs, three functions are commonly used - Power-law function,[23] Exponential function[24] and Sigmoid function.[25] In both power-law and exponential functions the stress concentration occurs at the portion (mostly near the surfaces) of the FGM structures where the composition change is continuous but rapid. By taking this into consideration sigmoid function has been proposed as a grading function for FGMs. It is found that sigmoid FGMs can significantly reduce the stress concentration factor near the surfaces.[25] The study regarding these grading functions in nanostructures can lead to new and sophisticated utilizations of functionally graded nanomaterials.

Most studies available in the literature employ linear mixture rule to obtain the effective mechanical properties of bulk FGM structures. However, several research works indicate that, a proper micromechanical model[26] should be implemented to accurately obtain the effective mechanical properties of FGMs.[27,28] One of the simplest micromechanical models to calculate effective mechanical properties of FGMs is Voigt model,[29] which is based on the assumption that all the phases within the structure experience same strain. On the other hand, Reuss model[30] assumes stress uniformity throughout the material. Another relevant micromechanical model is Hashin-Shtrikman bounds[31] which uses variational principle for heterogeneous linear elasticity. Tamura *et al.*[32] presented a model of determining effective elastic modulus that incorporates an empirical fitting parameter which relates the stress and strain in the constituent phases. Understanding the applicability of these models in FGM nanostructures and finding out which of these models is best suited for FGMs in nanoscale are of great importance for future applications of FGMs in nanoelectronics.

In the present study, a novel approach for modeling (section 2.2) FGM nanostructures using MD simulations is implemented to study the mechanical and vibrational properties of radially graded FGM nanowires. Various grading functions that govern the constituent distribution profiles of the NWs are considered. The effects of the grading functions on the mechanical and vibrational characteristics are investigated (section 3.1 and 3.4). Various explicit micromechanical models are applied in case of present FGM NWs to find out which of these models can be recommended to calculate the effective elastic modulus of radially graded FGM NWs (section 3.2). The effects of diameter of the FGM NWs on the mechanical properties have been analyzed considering different micromechanical models (section 3.3). Also, the results of vibrational studies are compared with both Euler-Bernoulli beam theory and He-Lilley theoretical model and their deviations are discussed (section 3.5). Finally, the origin of observed beat vibration in FGM NWs is investigated (section 3.6).

## 2. Methodology

### 2.1 Simulation domain and interatomic potential

In this study, mechanical properties and vibrational characteristics of radially graded FGM NW characterized by different grading functions have been analyzed through MD simulations. Two different MD simulations (tensile and vibrational test) are performed to get a detailed understanding of these properties and the effects of grading functions on them. We have selected *Cu* and *Ni* as the alloying constituents of radially graded FGM NWs. *Cu* and *Ni* are both widely used because of their excellent electronic, magnetic and catalytic properties.[33–35] Therefore, *Cu–Ni* alloy nanostructures exhibit advantageous properties such as high electronic conductivity, excellent magnetism and favorable chemical stability which could find potential applications in nano-electronics. Besides their vast applications in various industries, the reasons for choosing *Cu-Ni* alloy are that, below $1085^oC$, at all alloying percentages, *Cu-Ni* forms only a single *α*-phase and *Ni* atoms substitute *Cu* atoms randomly from copper's FCC lattice points in *α*-phase. Also, their nearly identical lattice constants (0.352 nm for *Ni* and 0.360 nm for *Cu*) results in a relatively insignificant lattice misfit and residual strain.

Cylindrical *Cu-Ni* FGM NWs have been used for both the tensile and vibrational test simulations as shown in Fig. 1. The dimensions and total number of atoms of the NWs used for both simulations are presented in Table 1.

**Table 1.** Dimensions and total number of *Cu* and *Ni* atoms in the FGM nanowires.

|  | Diameter, $D$ (nm) | Length, $L$ (nm) | Number of atoms |
|---|---|---|---|
| Tensile test | 5.76 | 60.66 | 147030 |
| Vibrational test | 3.78 | 20.70 | 22330 |

However, it is worth mentioning that, although the table presents the total number of atoms in the NWs, the individual numbers of *Cu* and *Ni* atoms vary depending on the grading function being used to construct the NWs which will be discussed in the next section. Both the NWs are arranged in a face-centered cubic (FCC) lattice structure modeled with *x*, *y* and *z* directions oriented along [100], [010], and [001] directions respectively. For the vibrational test, the dimensions of the NW with a slenderness ratio larger than 10 is selected under the assumption that the NW can be simplified as thin beam. All boundaries in the simulation domain are set free in this test. However, for the tensile test, the simulation domain boundaries in *x* and *y* directions are set free, while periodic boundary condition is applied along the loading direction (*z*-direction). The existence of free surfaces along the free boundaries will result in the relaxation motion of atoms near the curved surface of the NWs, which will then minimize the total potential energy of the system. At 0K, the lattice constant of *Cu* ($a_{Cu}$) is 3.6 Å and that of *Ni* ($a_{Ni}$) is 3.52 Å. So, the lattice misfit between *Cu* and *Ni* is ~2.2%. To build the FGM NWs with reasonably stable structure, an average lattice constant, $a_{FGM} = (a_{Cu}+a_{Ni})/2 = 3.56$ Å, is set as the lattice constant of *Cu* and *Ni* to arrange atoms in both simulation domains (tensile and vibrational tests).

The interactions between all the atoms within the simulation domain are described by the embedded atom method (EAM) potential.[36] In this method, the potential energy of an atom, *i*, is given by:

$$E_i = F_\alpha \left( \sum_{i \neq j} \rho_\beta(r_{ij}) \right) + \frac{1}{2} \sum_{i \neq j} \varphi_{\alpha\beta}(r_{ij}) \qquad (1)$$

where, $r_{ij}$ is the distance between atoms *i* and *j*, $\varphi_{\alpha\beta}$ is a pairwise potential function, $\rho_\beta$ is the contribution to electron charge density from atom *j* at the location of atom *i* and $F_\alpha$ is an embedding function that represents the energy required to place atom *i* into the electron cloud. *α* and *β* are the element types of atoms *i* and *j* respectively. The potential parameters developed by Onat and Durukano[37] are used in the present simulations. This potential can consistently reproduce a wide range of materials properties of *Cu*, *Ni* and *Cu-Ni* compounds, including enthalpy of mixing, melting points, defect formation energies and ground state lattice structures.[37,38] Extensive investigations of various properties of *Cu-Ni* alloys have been carried out using this potential in recent years.[39,40] Onat and Durukano[37] have demonstrated the capability of the potential to reproduce the phonons of *Cu-Ni* alloy FCC structures. This potential has also been used to study dislocation cross slip of *Cu* + (10, 22, 33, 68, 79, 90)% *Ni* FCC alloys.[40] Since *Cu-Ni* FGM structures consist of a wide range of continuously varying alloy compositions, and the aforementioned potential parameters accurately reproduce the properties of *Cu-Ni* alloys throughout alloying compositions. Hence, we selected these particular potential parameters for our present study.

## 2.2 Modelling method of FGM

FGM NWs can be produced by continuously changing the fraction of the constituent materials along radially outward direction. The function that determines this continuous change is known as the grading function. Our present approach of modeling FGM NWs is based on Solid Freeform Fabrication (SFF)[41] method, in which a solid model is subdivided into a number of sub-regions and each region is associated with a composition blending function. The modeling of radially graded *Cu-Ni* FGM NW consists of the following three steps: (1) a pure *Cu* NW of appropriate dimensions and lattice constant is prepared, (2) The NW is then subdivided into a number of annular cylindrical chunks of 1.78 Å (half of the lattice constant) thickness; the innermost chunk is a cylinder with a diameter of 3.56 Å (lattice constant), and (3) in each chunk, according to the grading functions, the *Cu* atoms are randomly replaced by *Ni* atoms, similar to the modelling method adopted by Mojumder *et al.*[42] Hence our present method of modeling FGM nanostructures combines the SFF[41] method of FGM fabrication and atomistic approach of modeling metal alloys[42] to model FGMs in nanoscale. It is worth mentioning that, this method of modeling FGM is especially applicable for FGM alloys where the lattice misfit between the constituents is negligible (e.g. copper-nickel, gold-platinum etc.). Also, if the NW is not sufficiently large (having dimensions of only a few lattice constants), then the grading will not be smooth enough to consider the NW functionally graded.

One of the most important factors that dictate the properties and applications of FGM is its grading function. FGM NWs graded by Power-law function (P-FGM), exponential function (E-FGM) and sigmoid function (S-FGM) are considered in this paper as they are the most commonly available in the scientific literature.[23–25] The mass fractions of *Ni* in *Cu*, *g(r)*, for P-FGM, E-FGM and S-FGM are presented in Table 2, where *p* is a function parameter, *r* is the radial distance from the center of the NW and *R* is the radius of the NW. By changing the function parameter *p*, a wide range of FGM NWs of different grading profiles can be generated. The distribution of mass fraction of *Ni* in *Cu* along the radial direction for different types of FGM NWs with a wide range of function parameters, *p* considered in this study are illustrated in Fig. 2. Top views of a few representative FGM NWs of different types generated by the present modeling scheme are shown in Fig. 3. The input structures used in the present study were generated using *LAMMPS Input Structure Generator for Functionally Graded Material (FGM) tool*[43] in nanoHUB (https://nanohub.org/tools/fgmbuilder).

**Table 2.** Mass fraction distribution of *Ni* in *Cu* for different types of FGM

| Type of FGM | Grading function | Mass fraction of *Ni* in *Cu*, g(r) |
|---|---|---|
| P-FGM | Power-law function | $g(r) = \left(\frac{r}{R}\right)^p$ |
| E-FGM | Exponential function | $g(r) = 1 - e^{-\left(\frac{r}{R}\right)^p}$ |
| S-FGM | Sigmoid function | $g(r) = \frac{1}{2}\left(2\frac{r}{R}\right)^p \text{ for } 0 \leq r \leq \frac{R}{2}$ <br> $g(r) = 1 - \frac{1}{2}\left(2\left(1-\frac{r}{R}\right)\right)^p \text{ for } \frac{R}{2} \leq r \leq R$ |

*2.3 Simulation procedure of tensile test*

At first, the geometries of the functionally graded NWs are minimized using a conjugate gradient minimization scheme. Then, the NW is sufficiently relaxed for 50 ps under NVE equilibration while keeping the temperature at 300 K using Langevin thermostat. Following the NVE equilibration, the pressure of the system is equilibrated by applying the isothermal-isobaric (NPT) ensemble in *z*-direction at 1 bar and a temperature of 300 K for 100 ps with a pressure damping parameter of 100 fs and a drag factor of 0.3. After that, the system is thermally equilibrated using the canonical (NVT) ensemble for 10 ps with a coupling constant of 100 fs. Finally, a uniaxial strain was applied along the *z*-direction at a constant strain rate of $10^9 \ s^{-1}$, which is considered to be a suitable strain rate for MD simulations due to the computational constraints of MD. The tensile tests are conducted with the NVT ensemble using a Nose–Hoover thermostat to keep a constant temperature of 300 K. This procedure of tensile test is similar to previous studies.[44,45] Each deformation simulation is carried out until the failure of the FGM nanowires occurred. We calculated the virial stress[46–48] using the formula:

$$\sigma_{virial}(\mathbf{r}) = \frac{1}{\Omega} \sum_i \left( -m_i \dot{\mathbf{u}}_i \otimes \dot{\mathbf{u}}_i + \frac{1}{2} \sum_{j \neq i} \mathbf{r}_{ij} \otimes \mathbf{f}_{ij} \right) \qquad (2)$$

where the summation is over all the atoms occupying the total volume, $\Omega$, $m_i$ is the mass of atom $i$, $\mathbf{u}_i$ is its displacement; $\dot{\mathbf{u}}_i = d\mathbf{u}_i/dt$ the velocity; $\mathbf{r}_{ij}$ is the position vector of atom $i$ with respect to atom $j$; $\otimes$ is the outer, dyadic or direct tensor product of two vectors; and $\mathbf{f}_{ij}$ is the force on atom $i$ due to the pair interaction with atom $j$. Elastic modulus, $E$, is calculated using the gradient of the elastic region of the stress-strain curve.

### *2.4 Simulation procedure of vibrational test*

In case of vibrational tests, the NW is first relaxed to a minimum energy state using conjugate gradient energy minimization. After that, the NWs are relaxed under NVE equilibration with Langevin thermostat turned on to keep the temperature fixed at 300 K for 50 ps. After that, the system is thermally equilibrated using the canonical (NVT) ensemble for 100 ps with a temperature damping parameter 10 fs. A Nose–Hoover thermostat was employed to keep the temperature constant at 300 K. At this point of the simulation, three lattice constants length ($3a_{FGM}$=10.68 Å) from the two ends of the NWs are kept rigid to simulate the clamped-clamped boundary condition. Then an initial sinusoidal transverse velocity field is applied to the NW. The initial velocity field can be expressed as:

$$u'(z) = \lambda \sin(kz) \qquad (3)$$

where, $\lambda$ is the initial velocity amplitude, $k$ is equal to $\pi/L'$, $u(z)$ is the displacement from the straight nanowire position in $x$ direction and $u'(z)$ is the corresponding velocity. In the present study, the $\lambda$ equals 1 Å/ps. After the initial velocity excitation, the vibration of the NW is achieved under constant NVE ensemble for 10 ns. This procedure of vibrational test is similar to previous studies.[21] As the total energy is preserved during vibration due to NVE ensemble, the lost potential energy is converted into kinetic energy. Thus by following the method adopted by Zhan and Gu,[21] from the time history of external energy (*EE*), the first order natural frequency of the NWs can be calculated through the application of fast Fourier transform (FFT), which gives a fast computation of discrete Fourier transform (DFT).

During the oscillations of the NWs, there is a cyclic conversion between their potential energy and kinetic energy. However, the maximum external potential energy, $E_{ext}$, decreases by $\Delta E_{ext}$ in each oscillation cycle due to damping. The quality factor $Q$, is thus defined as:

$$Q = \frac{2\pi E_{ext}}{\Delta E_{ext}} \quad (4)$$

During the latter part of the oscillations (where the effect of beat phenomenon has diminished), the *Q*-factor can be considered to remain fairly constant. Therefore, for those parts of the oscillations, at the end of any *n* cycles, the maximum external potential energy $E_n$ is related to the maximum potential energy prior to those *n* oscillations, $E_{ext}$, by:

$$E_n = E_{ext}\left(1 - \frac{2\pi}{Q}\right)^n \quad (5)$$

Thus, using equation (5), the quality factors of the different FGM NWs have been calculated and compared in this study. This method has been previously adopted by Jiang *et al.* to calculate the quality factor of CNT beams.[49]

To calculate the atom propagation, the equations of motion are integrated using a velocity verlet algorithm with a timestep of 1 fs for both tensile tests and vibrational tests. All the simulations are performed using LAMMPS,[50] and visualization is done using Ovito.[51]

In our present study we have performed 20 tensile test simulations (147030 atoms) and 18 vibration test simulations (22330 atoms). In order to investigate diameter effect, 9 tensile test simulations (197060 atoms) have been conducted. Also, for comparison with theoretical results, we performed another 18 tensile test simulations (22330 atoms) of the NWs used in vibration test. For the vibration test simulations, we made specific use of the LAMMPS GPU (Graphics Processing Unit) package,[52] and for each simulation it took approximately 12 hours wall time using 1 GPU per node. We performed tensile test simulations using 24 processors, and each simulation required about 3 hours wall time. On similar configuration, the tensile tests of NWs used in vibration test and in the study of diameter effect took approximately 0.5 and 4 hours, respectively.

## *3. Results and Discussion*

### *3.1 Effects of grading function of FGM on mechanical characteristics*

In this study, three different types of grading functions have been considered and the function parameter, $p$ has been varied to understand its modulating effects on the mechanical characteristics of FGM NWs. Considering a wide range of function parameters, the stress-strain curves of P-FGM, E-FGM and S-FGM NWs are shown, along with pure *Cu* and pure *Ni* NWs, in Fig. 4(a), (b) and (c) respectively.

From Fig. 4(a), it can be seen that, the value of strains at the failure points are around 11-13% for P-FGM NWs. After failure the presence of many serrations in the stress−strain curves, indicates that the failures of P-FGM NWs are ductile in nature. The failure modes of E-FGM and S-FGM NWs are also ductile as shown in Fig. 4(b) and (c). This mode of failure is expected as both *Cu* and *Ni* are fairly ductile materials. The ranges of failure strain values of E-FGM and S-FGM NWs are also similar to those of P-FGM NWs. Elastic modulus (*E*) of the NWs are calculated from the elastic region of the stress-strain curves. The variation of *E* with *p* for P-FGM, E-FGM and S-FGM NWs are illustrated in Fig. 5(a), (b) and (c) respectively. From the figures, it is apparent that $E$ decreases with increasing value of $p$ in case of P-FGM and E-FGM NWs. Also, for S-FGM NWs, $E$ increases as the value of $p$ increases. However, this increase in $E$ with $p$ for S-FGM NWs is very small compared to that of P-FGM and E-FGM NWs. Another important mechanical property of NWs is ultimate tensile strength ($UTS$). The variation of $UTS$ with $p$ for the considered types of FGM NWs are shown in Fig. 6. In case of P-FGM and E-FGM NWs, with the increase of $p$ from 0.1 to 10, the $UTS$ decreases. The trend is opposite in case of S-FGM NWs. Similar to $E$, the increase of $UTS$ with increasing $p$ for S-FGM NWs is also small compared to P-FGM and E-FGM NWs.

The variation trend of $E$ and $UTS$ with respect to $p$ can be explained by considering total alloying percentages of $Cu$ and $Ni$ and the profile of distribution. Elastic modulus of bulk $Ni$ ($E_{Ni}$ = 207 GPa) is greater than that of bulk $Cu$ ($E_{Cu}$ = 128 GPa). Also, ultimate tensile strength of $Ni$ ($UTS_{Ni}$ = 317 MPa) is significantly larger than that of $Cu$ ($UTS_{Cu}$ = 200 MPa). Hence for FGM NWs with high $Ni$ constituents, the values of $E$ and $UTS$ will be higher compared to that of FGM NWs with low $Ni$ constituents. In Fig. 2(a), (b) and (c), the area under the profile curves approximately represent the total fraction of $Ni$ constituent in the NWs. It is clear from Fig. 2(a) and (b) that, total percentage of $Ni$ constituent increases as the value of $p$ decreases from 10 to 0.1 in case of P-FGM and E-FGM NWs. But in Fig. 2(c), the area under the curve remains constant for all values of $p$. Therefore, the total $Ni$ fraction remains almost same regardless of the value of function parameter $p$ for S-FGM. This is the reason why the value of $E$ and $UTS$ changes slightly with the change of $p$, as mentioned earlier. Even though the increase of $E$ and $UTS$ with the increase of $p$ is small, there is an increase nonetheless. This slight increase can be attributed to the distribution profile of S-FGM NWs. From Fig. 2(c), it is apparent that, when $p = 1$, the distribution is linear, which means that *Ni* percentage increases gradually and *Cu-Ni* alloy exists throughout the NW. As the value of *p* increases, the distribution profiles change and at *p* = 10, there exists a sudden change of *Ni* percentage. Thus, for *p* = 10, *Cu-Ni* alloy is only present in the region where the sudden change of *Ni* percentage occurs. As mentioned earlier, *Cu* and *Ni* atoms have lattice misfit and therefore when they are in an alloy together there are a large number of point defects after equilibration relative to pure structures. To calculate the number of point defects in the S-FGM NWs after equilibration, Wigner-Seitz defect analysis are conducted. As there is no point defect source in the FGM NWs, the numbers of vacancies and interstitials created after the equilibration are always same and they are known as vacancy-interstitial pairs. The numbers of vacancy-interstitial pairs in S-FGM NWs with different values of *p* calculated from Wigner-Seitz defect analysis are tabulated in Table 3. It is clear that as the value of *p* increases from *p* = 1 to *p* = 10, the number of point defects also increases. These point defects decrease the strength of the NWs. As a result, large alloyed portion of S-FGM, *p* = 1 compared to small alloyed portion of S-FGM, *p* = 10 contributes to the slightly higher *E* and *UTS* of S-FGM, *p* = 10 than that of S-FGM, *p* = 1 by accommodating more point defects. This result shows that, the grading function in case of S-FGM NWs affects the mechanical properties even though the constituent weight fraction is nearly same. This provides a background for further research on the comparison of FGMs and alloys with same constituent weight fractions. It is worth mentioning that the values of *E* and *UTS* of FGM NWs considered in this study are significantly different from their bulk counterparts. As the size of the NW decreases, the surface area to volume ratio increases and resulting higher surface energy begins to influence the mechanical properties of the NW.[53] The characteristics of the values of *E* and *UTS* increase or decrease due to the surface energy effect, which depends on two competing factors such as atomic coordination and electron redistribution.[54] In the present study, the values of *E* are smaller than bulk values while *UTS* values are larger than their bulk counterparts.

**Table 3.** Number of vacancy-interstitial pairs in S-FGM NWs for different values of *p* obtained from Wigner-Seitz defect analysis.

| Function parameter, *p* | Vacancy-interstitial pairs |
|:---:|:---:|
| 1 | 10194 |
| 2 | 8422 |
| 4 | 8279 |
| 10 | 7505 |

*3.2 Applicability of different micromechanical models and bounds on elastic modulus*

There are different micromechanical models available in the literature to predict the effective elastic modulus of alloys and composites based on weighted contributions from their constituents. These models are briefly introduced in section 1. *Cu-Ni* FGM NWs considered in the present study are essentially alloys of *Cu* and *Ni,* where the alloying fraction varies along the radial direction. In order to understand the applicability of these models in case of nanoscale FGM structures and to find out which of these models is best suited for FGM NWs, the elastic modulus (*E*) obtained from the present MD simulations are compared with those obtained from the aforementioned models. The expression of effective elastic modulus for different micromechanical models are presented in Table 4 where, $g(r)$ is the mass fraction of *Ni* in *Cu*, $E_{Cu}$ and $E_{Ni}$ are the elastic moduli of pure *Cu* and pure *Ni* NWs respectively and $q_T$ is the empirical fitting parameter for Tamura model. The NWs of pure *Cu* and pure *Ni* are of the same dimensions as the FGM NWs considered in the present study and *E* and *UTS* values are obtained from MD simulations. In this work, for calculations involving Hashin-Shtrikman bounds, we estimated the value of the Poisson's ratio, *v*, at any part of the NWs to be constant at 0.3. This value of *v* is chosen to be constant to simplify the expression of effective $E$ in case of Hashin-Shtrikman bounds. Also, from the MD simulation results, the Poisson's ratio of different FGM NWs with different grading functions are calculated and their values are found to be scattered around 0.3. The empirical fitting parameters, $q_T$ required in Tamura model for P-FGM, E-FGM and S-FGM are calculated by minimizing the root mean square relative deviation (RMSRD) of Tamura model from MD results. For P-FGM and E-FGM NWs, the values of $q_T$ are -199 GPa and -72.2 GPa respectively (both are negative). On the other hand, for S-FGM the value of $q_T$ is 83.7 GPa.

**Table 4.** Expression of effective $E$ in various micromechanical models with their root mean squared (RMS) relative deviations with MD results.

| Model | Expression of effective elastic modulus | RMS relative deviation (%) | | |
|---|---|---|---|---|
| | | P-FGM | E-FGM | S-FGM |
| Voigt Model[29] | $E(r) = g(r)E_{Ni} + (1 - g(r))E_{Cu}$ | 14.3 | 13.6 | 15.6 |
| Hashin-Shtrikman upper bound[31] | $E(r) = E_{Ni} + \dfrac{1 - g(r)}{\dfrac{1}{(E_{Cu} - E_{Ni})} + \dfrac{(1+v)g(r)}{3E_{Ni}(1-v)}}$ | 9.9 | 6.3 | 12.2 |
| Hashin-Shtrikman lower bound[31] | $E(r) = E_{Cu} + \dfrac{g(r)}{\dfrac{1}{(E_{Ni} - E_{Cu})} + \dfrac{(1+v)(1-g(r))}{3E_{Cu}(1-v)}}$ | 8.5 | 4.9 | 11.2 |
| Reuss Model[30] | $E(r) = \left(\dfrac{g(r)}{E_{Ni}} + \dfrac{1 - g(r)}{E_{Cu}}\right)^{-1}$ | 6.5 | 3.0 | 9.7 |
| Tamura Model[32] | $E(r) = \dfrac{(1 - g(r))E_{Cu}(q_T - E_{Ni}) + g(r)E_{Ni}(q_T - E_{Cu})}{(1 - g(r))(q_T - E_{Ni}) + g(r)(q_T - E_{Cu})}$ | 3.3 | 2.8 | 3 |

As there is no variation of *Cu* and *Ni* fraction along the longitudinal direction and the variation only exists along the radial direction, to obtain an average value of $E$, an integral along the radial direction is to be performed. So, the average values of $E$ obtained from the considered models can be written as:

$$\bar{E} = \frac{1}{\pi R^2} \int_0^R E(r) 2\pi r \, dr \tag{6}$$

The variation of $\bar{E}$ obtained from different models with $p$, for different types of FGM NWs have been compared with results obtained from MD simulations in Fig. 5. To compare the micromechanical models with MD results quantitatively, we used the term root mean square relative deviation (RMSRD), which can be expressed as:

$$RMSRD = \sqrt{\frac{1}{N_p} \sum_p \left( \frac{\bar{E} - E_{MD}}{E_{MD}} \right)^2} \tag{7}$$

Here, $E_{MD}$ is the elastic modulus obtained from the MD simulations and $N_p$ is the number of function parameters for FGM NWs of any specific grading function. The RMSRD of different models from MD results have been tabulated in Table 4. It is illustrated in Fig. 5(a) that, apart from Tamura model, all other models overestimate the values of $E$ in case of P-FGM. One important observation from the Table 4 is that, among all the considered models, Voigt model is the least reliable and yields a maximum RMSRD of 14.3 %. Due to the simplicity of Voigt model, it is often used to determine the effective $E$ of bulk FGM structures. However, for nanoscale FGM structures such as FGM NWs, Voigt model should be avoided. Figure 5(a) along with Table 4 also shows that, among the non-empirical micromechanical models considered in the present study, Reuss model provides the most accuracy in case of P-FGM NWs with 6.5 % RMSRD. From Fig. 5(b) it can be observed that, $E$ values obtained from Voigt model are higher than MD results for E-FGM NWs. Table 4 points out that, both Reuss and Tamura models are equally reliable with RMSRD of 3 % and 2.8 %, respectively for E-FGM NWs. As Reuss model is non-empirical, it is simpler and easier to implement compared to empirical Tamura model with almost same accuracy in case of E-FGM NWs. Figure 5(c) shows that, Voigt, Ruess and Hashin-Shtrikman bounds overestimate the values of $E$ in case of S-FGM. And similar to P-FGM, Reuss model offers better accuracy (9.7 % RMSRD) compared to other non-empirical models. In Fig. 5(c) it is illustrated that, the Tamura model suggests no effects of $p$ on $E$ in case of S-FGM. As the value of $p$ increases, all non-empirical models for S-FGM NWs converge. Also, when $p > 10$, values of $E$ for all the models (except empirical Tamura model) become very close. From Fig. 3, it is clear that S-FGM $p = 10$, closely resembles core-shell NW and in that case, any of the non-empirical models considered in the present study will yield similar effective $E$ values.

### *3.3 Effects of diameter of FGM on mechanical properties*

To understand the effects of FGM NW diameter on mechanical properties, we conducted a number of tensile test simulations on P-FGM NWs of 6.66 nm diameter while maintaining the same length (60.66 nm) as of the previously considered NWs. The variation of elastic modulus, $E$ with $p$ for P-FGM NWs of both diameters are compared in Fig. 7(a). As we see from Fig. 7(a), with the increase of diameter, the values of $E$ increase for all the values of $p$. This trend is consistent with the previous numerical[54] and experimental studies[53] on pure FCC nanowires. Similar to pure $Cu$[54] and $Au$[53] NWs, the increase of $E$ with the increase of diameter can be attributed to nonlinear elastic deformation of NW core and NW surface effect. Also the increase in NW diameter results in similar increase of $E$ for different values of $p$. This allude to the fact that, the effects of diameter on the value of $E$ are independent of the grading function parameter. The variations of *UTS* with $p$, for both diameters considered are shown in Fig. 7(b). The values of *UTS* of the P-FGM NWs of 6.66 nm diameter are larger than those of 5.76 nm diameter NWs for all the values of *p*. This increase of *UTS* due to diameter increase is roughly same across all values of *p*. Therefore, the effects of diameter on the value of *UTS* are also independent of the grading function parameter. Our present comparison between P-FGM NWs of two different diameters shows that, the effects of diameter on the mechanical properties of FGM NWs are similar to that of pure metallic NWs.

From the previous section (section 3.2), we concluded that, among non-empirical micromechanical models, Ruess model is the most accurate model to describe the elastic modulus of FGM NWs. Also, it was found that, Voigt model is the most unreliable among non-empirical micromechanical models. With the increase of NW diameter from 5.76 nm to 6.66 nm, these observations remain similar. The RMSRD of MD results from different micromechanical models for both diameters are illustrated in Fig. 8. It is clear that regardless of change in NW diameter, Reuss model remains the most reliable non-empirical micromechanical model. Although the RMSRD from Reuss model decreases with the increase of NW diameter, this decrease is less than 2%. On the other hand, Fig. 8 shows that, the accuracy of empirical Tamura model decreases with diameter increase and this decrease is also not that significant (less than 1%). Fig. 8 also illustrates that, although diameters of the P-FGM NWs are increased by more than 15 %, the RMSRD for all the considered micromechanical models do not vary more than 5 %. Therefore, there is no significant effect of diameter on the applicability of micromechanical models in case of P-FGM NWs. It is worth mentioning that, in considering diameter effect, we only investigated P-FGM NWs in our present study as E-FGM and S-FGM NWs are expected to show similar diameter effects on the mechanical properties.

*3.4 Effects of grading function of FGM on vibrational characteristics*

The vibrational tests of different types of FGM NWs with end conditions as clamped-clamped (C-C) have been carried out by MD simulations. Two important vibrational properties are calculated in the present study, which are first order natural frequency and quality factor ($Q$-factor). The modulation effects of function parameter, $p$, on these properties are also studied. Variation of natural frequency with $p$ for P-FGM, E-FGM and S-FGM are illustrated in Fig. 9(a), (b) and (c), respectively. The FGM NWs considered in the study exhibit beat phenomenon. Therefore, the natural frequency calculated in the study is an average of two very close frequencies which are obtained from FFT calculation. In case of both P-FGM and E-FGM NWs, increasing value of $p$ results in decreasing natural frequency. As shown in Fig. 9(a), the maximum and minimum natural frequencies of P-FGM NWs are 34.84 GHz ($p = 0.1$) and 24.81 GHz ($p = 10$) respectively. Also, the maximum and minimum values of natural frequencies in case of E-FGM are 30.74 GHz and 24.62 GHz, respectively. The values of natural frequencies obtained from the present MD study are comparable with the values reported by Zhan and Gu.[21] Thus, by changing the value of $p$, a wide range of natural frequencies can be obtained from P-FGM and E-FGM NWs. It can be seen from Fig. 9(c), for S-FGM, the value of $p$ has little effect on natural frequency. This shows that, constituent fraction plays an important role in determining the natural frequency as S-FGM NWs have a constant *Cu* and *Ni* fraction. In case of P-FGM and E-FGM, the *Ni* percentage decreases with increasing value of $p$, which means that, natural frequency of *Cu-Ni* FGM NWs increases with increasing *Ni* constituent. This lowering of natural frequency with increase in *Ni* constituent can be attributed to the fact that the mass of *Ni* is lower than the mass of *Cu*. Therefore, NWs with high *Ni* fraction will be lighter than NWs with high *Cu* fraction and the lighter NWs are expected to exhibit high natural frequencies.

Variations of $Q$-factor with $p$ for different types of FGM NWs are illustrated in Fig. 10. In present study, we neglected any extrinsic damping effect on $Q$-factor by assuming the FGM NWs oscillate in vacuum and by making the support atoms on both ends rigid. The values of $Q$-factor ranges from $0.9 \times 10^4$ to $10.46 \times 10^4$ for P-FGM NWs, $0.8 \times 10^4$ to $3.17 \times 10^4$ for E-FGM NWs and $2.19 \times 10^4$ to $3.36 \times 10^4$ for S-FGM. These values of $Q$-factor are comparable to the values reported in previous studies.[21,49] For all types of FGM NWs, the $Q$-factor decreases with increasing values of $p$. So, a wide range of $Q$-factor can be obtained by changing the value of function parameter, $p$. As mentioned earlier, there are no extrinsic damping effects in the present vibrational test simulations which implies that, the $Q$-factor only depends on intrinsic damping factors. Two of the intrinsic factors widely reported in literature that influence $Q$-factor of NWs are thermoelastic damping[55] and atomic friction.[56] In the present study, the vibrational test simulations are carried out at 300K and during the vibration, due to initial velocity excitation the temperature of the NWs oscillates about mean temperatures of 330K to 340K. This indicates that thermoelastic damping is not that significant as the temperature at vibration for FGM NWs of different grading functions and function parameter are not drastically different. However, as the value of $p$ increases, the atomic configuration and constituent distribution profile of the FGM NWs change significantly, which may lead to higher atomic friction causing the value of $Q$-factor to decrease. The effects of atomic configuration and constituent distribution profile on atomic friction can be a topic of further study.

It can be observed from Fig. 3 that for higher values of $p$, the material distribution of the S-FGM NW approaches to that of a core-shell NW. Fig 10 (c) presents that the quality factor of an S-FGM nanowire with $p = 1$ is about 53% higher than that for $p = 10$. However, the atomic weight fractions are roughly same in these two NWs. These results show that, FGM NWs can potentially have higher quality factors than core shell NWs with similar atomic weight fraction.

### *3.5 Comparison of natural frequency with classical theories*

The NWs used for the vibrational test simulations can be considered as thin beams with C-C end condition. Hence by using Euler-Bernoulli beam theory (EBBT), the natural frequencies of the FGM NWs can be calculated with a certain accuracy. According to EBBT, the natural frequency of both ends clamped beam can be expressed as:

$$f_n = \frac{\omega_n}{2\pi L^2}\sqrt{\frac{EI}{\rho A}} \qquad (8)$$

where, $\omega_n$ is the eigenvalue obtained from the characteristic equation of $\cos(\sqrt{\omega_n})\cosh(\sqrt{\omega_n}) = 1$, $EI$ is flexural rigidity, $\rho$ is density and $A$ is cross-sectional area of the NW. Here, to avoid complications, the value of $\rho$ is taken as 8960 kg/$m^3$ as the densities of *Ni* and *Cu* are nearly identical. Also, the values of $E$ for these particular FGM NWs have been obtained by performing separate tensile test MD simulations. As first order vibration mode is of interest in present study, the value of $\omega_n$ used for theoretical calculations is 22.4.

The variation of natural frequency obtained from EBBT with *p* for P-FGM, E-FGM and S-FGM are shown in Fig. 9(a), (b) and (c), respectively. Also, the EBBT natural frequencies are compared with natural frequencies obtained from MD simulations in these figures. Similar to MD natural frequencies, the values of EBBT natural frequencies decrease with increasing *p* in case of P-FGM and E-FGM NWs. As seen in Fig. 9(c), for S-FGM NWs, the EBBT natural frequencies remain almost same for all values of *p*. However, the EBBT natural frequencies are considerably smaller than that obtained from the corresponding MD simulations. For all types of FGM NWs considered in this study, the natural frequencies obtained from MD simulation are found to be 5-17% larger than EBBT natural frequencies. This deviation of natural frequencies between MD values and EBBT predictions agree with the work of He and Lilley.[57]

One of the limitations of EBBT is that, it does not take the influence of surface stress on natural frequency into consideration. In order to eliminate this limitation, the theoretical model developed by He and Lilley[57] incorporated generalized Young-Laplace equation into EBBT. According to He-Lilley theoretical model (HLTM), the natural frequency of both ends clamped beam can be expressed as:

$$f_{ns} = f_n\sqrt{0.025\eta + 1} \qquad (9)$$

where, $f_n$ is the natural frequency obtained from EBBT and $\eta$ is a non-dimensional surface effect factor. For cylindrical NW, this factor is defined as:

$$\eta = \frac{2\tau D L^2}{(EI)^*}$$

Here, $D$ is the diameter of the NW, $\tau$ is the surface stress and $(EI)^*$ is the effective flexural rigidity considering surface elasticity. As previously mentioned, the values of $E$ are obtained by performing separate tensile test MD simulations that take surface elasticity into consideration. Therefore, we used these values of $E$ directly in our calculation of $(EI)^*$. For the sake of simplicity, the value of $\tau$ is taken as 2.502 N/m, which is the surface stress of *Ni* calculated at 298 K by Jiang *et al.*[58] Natural frequencies obtained from HLTM are compared with MD results in Fig. 11 for different types of FGM NWs. The variational trend of HLTM natural frequency with $p$ is similar to that of our present MD results for all grading functions considered in the present study. It is clear from the figures that, HLTM provides a better prediction of MD natural frequencies compared to EBBT. This result demonstrates that, positive surface stress increases the natural frequency of FGM NWs with C-C end condition and this effect is reflected in MD simulation results as well.

### *3.6 Beat Phenomenon*

In all vibrational tests conducted in the present study, beat vibration is identified. To illustrate the beat vibration, the time history of *EE* in a representative case of E-FGM, $p = 5$ is shown in Fig. 11 (a). At the beginning of the vibration, the amplitude of *EE* decays significantly. The *EE* amplitude shows a periodic pulsation pattern after the initial decay of amplitude. This *EE* amplitude trend is usually generated by two vibrations of very close frequencies, which is called beat vibration. As mentioned earlier, the initial velocity excitation has been given along only *x*-direction. However, it is clear from Fig. 11(c) and 11(d) that after a few oscillations, the NW (E-FGM, $p = 5$) not only vibrates along *x*-direction but also along *y*-direction. Gil-Santos *et al*.[59] reported that when the cross-sectional symmetry of Si NWs is broken, two vibration in orthogonal planes originate from a single planer vibration. In FGM NWs examined in this study, the cross-sections of the NWs are not symmetric in terms of material distribution; as during modeling of FGM NWs, *Ni* atoms randomly replace *Cu* atoms. This asymmetry in FGM NWs results in minor difference of natural frequencies along these two directions. The existence of two different vibrational frequencies can also be observed in one-sided FFT spectrum as shown in Fig. 11(b). Thus a single initial excitation initiates two different vibrations of the FGM NW. This is because the FGM NWs possess two orthogonal elementary atomic orientations. When the initial excitation is given along any direction other than these two orthogonal elementary directions, the initial excitation decomposes, generating two vibrations of nearly equal frequencies. This process of decomposition of initial excitation has been suggested by Zhan *et al*.[60] in case of [1 1 0] oriented *Ag* NW.

Single planer vibration of NWs is assumed in most nano electro-mechanical system (NEMS) applications. However non-planer oscillation of NWs can also be exploited for a wide range of applications such as mass sensing and stiffness spectroscopy.[59] It is interesting that FGM NWs, due to its asymmetrical cross-section will always exhibit non-planer vibration which leads to beat vibration. Therefore, FGM NWs of different types of grading functions can have numerous applications in NEMS that operate in the non-planer vibration regime.

## 4. Conclusions

We used an atomistic scale modeling approach for generating *Cu-Ni* functionally graded materials (FGM) NWs to study their mechanical and vibrational characteristics through MD simulation. Three grading functions governing the distribution of *Ni* in *Cu* (Power-law, exponential and sigmoid functions) are considered in the present study. To perform tensile test and vibrational test simulations, we constructed a wide range of FGM NWs by changing the function parameter *p*. The tensile test results show that, by changing the value of *p*, the Elastic modulus (*E*) and Ultimate tensile strength (*UTS*) can be modulated, especially in cases of P-FGM and E-FGM NWs. It is also found that, in case of S-FGM NWs, *p* has little influence on *E* and *UTS*. The *E* values obtained from Voigt, Ruess, Hashin-Shtrikman and Tamura micromechanical models are compared with the present MD simulation results. Voigt model exhibits the largest discrepancy with MD results among all the considered models. Reuss model displays relatively good accuracy among non-empirical models, especially in case of E-FGM NWs. With proper empirical fitting parameter, Tamura model can be a great tool in predicting effective *E* of FGM NWs. The effects of diameter on the FGM NWs mechanical properties are similar to that of pure metal NWs. Natural frequencies and *Q*-factors of different types of FGM NWs are obtained by performing vibrational test simulations. Function parameter can be changed to regulate the natural frequencies of P-FGM and E-FGM NWs. *Q*-factor is also affected by function parameter but further studies are required to understand the dependency of *Q*-factor on atomic configuration and constituent distribution profile of FGM NWs. It is also observed that FGM NWs can potentially have higher *Q*-factors than core shell NWs with similar atomic weight fractions. Theoretical values of natural frequencies of the considered FGM NWs are calculated using Euler-Bernoulli beam theory (EBBT) and He-Lilley theoretical model (HLTM). They follow the same variational trend with *p* as the natural frequencies obtained from MD results. The deviation of MD results from EBBT predictions stems from the positive surface stress of the FGM NWs. Since HLTM incorporates surface stress into its calculation of natural frequencies, this model can be considered more appropriate for FGM NWs. Beat vibrations of the FGM NWs are identified in the present study. We found that the beat vibrations can originate from the decomposition of initial excitation due to the existence of two orthogonal elementary orientations in FGM NWs. This can have significant applications in the non-planer vibration regime of NEMS. The results demonstrated in this work provide a new perspective on the mechanics of FGM NWs and would guide experimental synthesis of these NWs to obtain the desired mechanical and vibrational properties. As an advanced material with numerous potential applications, various FGM nanostructures and their properties should be studied extensively. The present modeling scheme along with our developed nanoHUB tool[43] open up a new avenue in the investigation of FGMs in nanoscale through molecular dynamics simulations.

**Acknowledgement**

The authors would like to acknowledge Multiscale Mechanical Modeling and Research Networks (MMMRN) for their technical assistance to carry out the research. M.M.I acknowledges the support from Wayne State University startup funds.

**Conflict of Interest**

There are no conflicts to declare.

**Author Contribution Statement**

**Mahmudul Islam:** Methodology, Validation, Formal analysis, Data Curation, Visualization, Writing – Original Draft. **Md Shajedul Hoque Thakur:** Methodology, Validation, Software, Data Curation, Visualization, Writing-Review & Editing. **Satyajit Mojumder:** Conceptualization, Methodology, Resources, Writing - Review & Editing, Project administration. **Abdullah Al Amin:** Writing - Review & Editing. **Md Mahbubul Islam:** Supervision, Writing - Review & Editing, Funding acquisition.

| | **List of Figure Captions** |
|---|---|
| **Figure 1** | Schematic of **(a)** FGM NW used for tensile test and **(b)** clamped-clamped FGM NW used for vibrational test. |
| **Figure 2** | Distribution of mass fraction of *Ni* in *Cu* along the radial direction for **(a)** P-FGM, **(b)** E-FGM and **(c)** S-FGM NWs for different values of function parameter *p*. |
| **Figure 3** | Top views of representative FGM NWs of different distributions for different values of *p*. |
| **Figure 4** | Stress-strain curves of **(a)** P-FGM, **(b)** E-FGM and **(c)** S-FGM NWs for different values of function parameter *p*. |
| **Figure 5** | Variation of Elastic moduli of **(a)** P-FGM, **(b)** E-FGM and **(c)** S-FGM NWs for different values of function parameter, *p*, and comparisons with the Voigt model, Reuss Model, Hashin–Shtrikman (HS) upper and lower bounds and the Tamura model with appropriate values of $q_T$. |
| **Figure 6** | Variation of Ultimate tensile stress of P-FGM, E-FGM and S-FGM NWs for different values of function parameter, *p*. |
| **Figure 7** | Effect of P-FGM NW diameter on **(a)** Elastic Modulus and **(b)** Ultimate Tensile Strength, at different values of function parameter, *p*. |
| **Figure 8** | Effect of P-FGM NW diameter on the RMS relative deviations of MD results from different micromechanical models. |
| **Figure 9** | Variation of Natural Frequency of **(a)** P-FGM, **(b)** E-FGM and **(c)** S-FGM NWs for different values of function parameter, *p*, and comparison with the predictions of Euler-Barnoulli Beam Theory and He-Lilley theoretical model. |
| **Figure 10** | Variation of Q-factor of **(a)** P-FGM, **(b)** E-FGM and **(c)** S-FGM NWs for different values of function parameter, *p*. |
| **Figure 11** | **(a)** Time history EE during vibration, **(b)** one-sided FFT power spectrum of EE in frequency domain, atomic configuration at 900 ps to visualize vibration along **(c)** x-direction and **(d)** y-direction, *f*or E-FGM, *p* = 5 NW. |

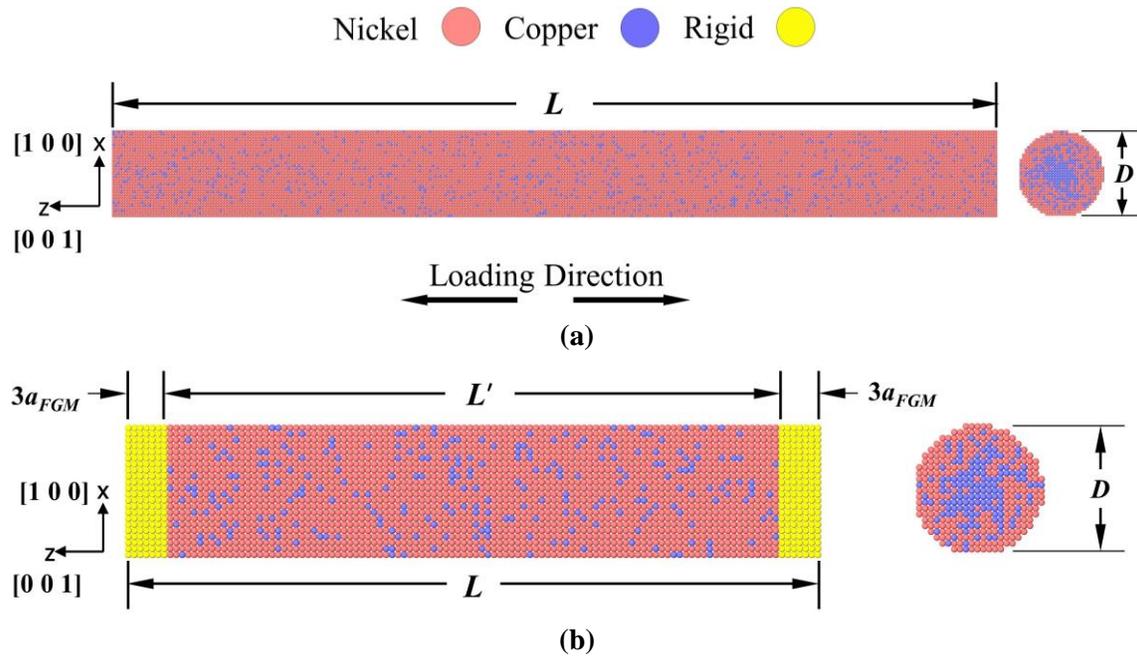

**Figure 1.** Schematic of (**a**) FGM NW used for tensile test and (**b**) clamped-clamped FGM NW used for vibrational test.

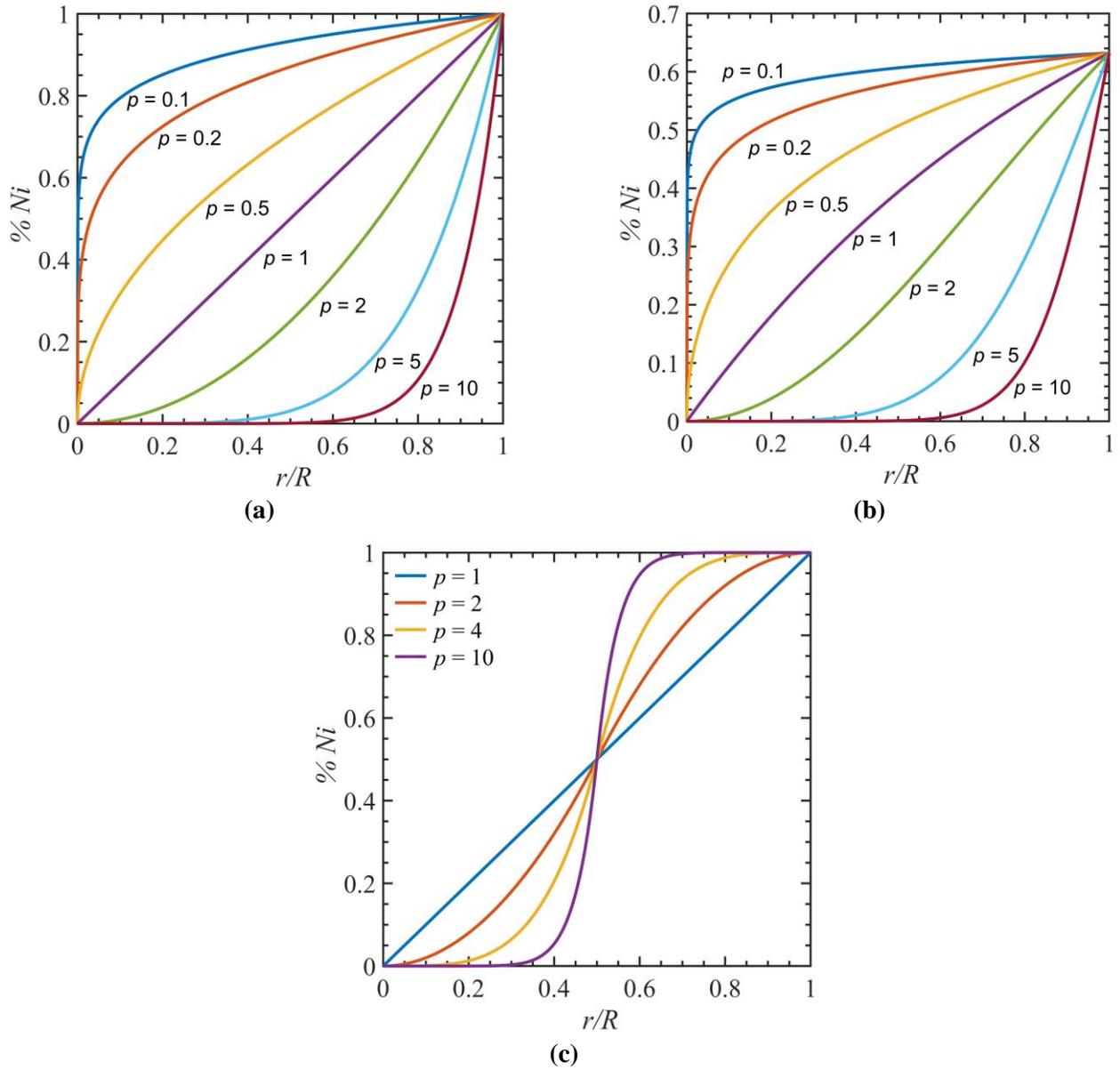

**Figure 2.** Distribution of mass fraction of *Ni* in *Cu* along the radial direction for **(a)** P-FGM, **(b)** E-FGM and **(c)** S-FGM NWs for different values of function parameter *p*.

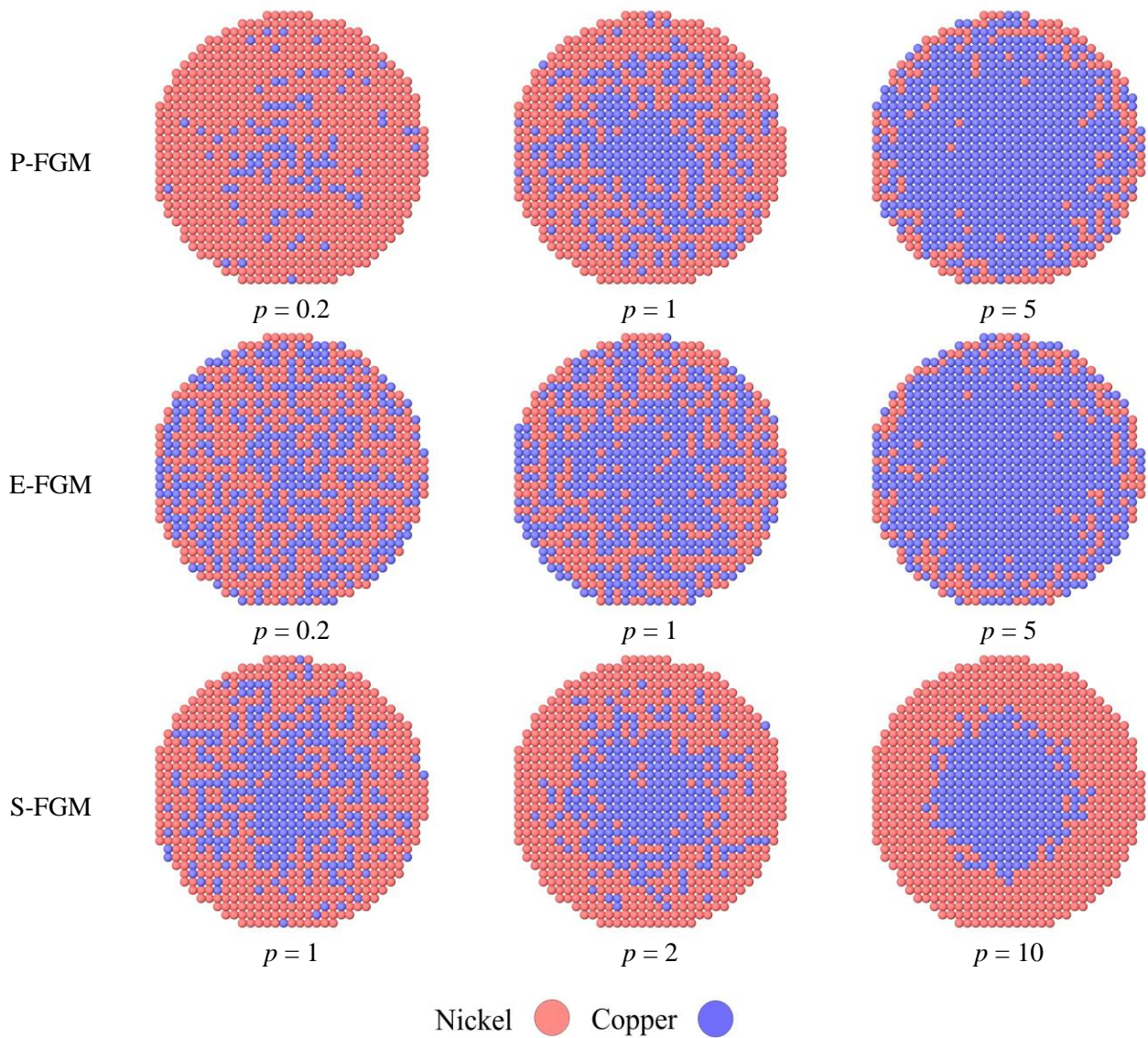

**Figure 3.** Top views of representative FGM NWs of different distributions for different values of *p*.

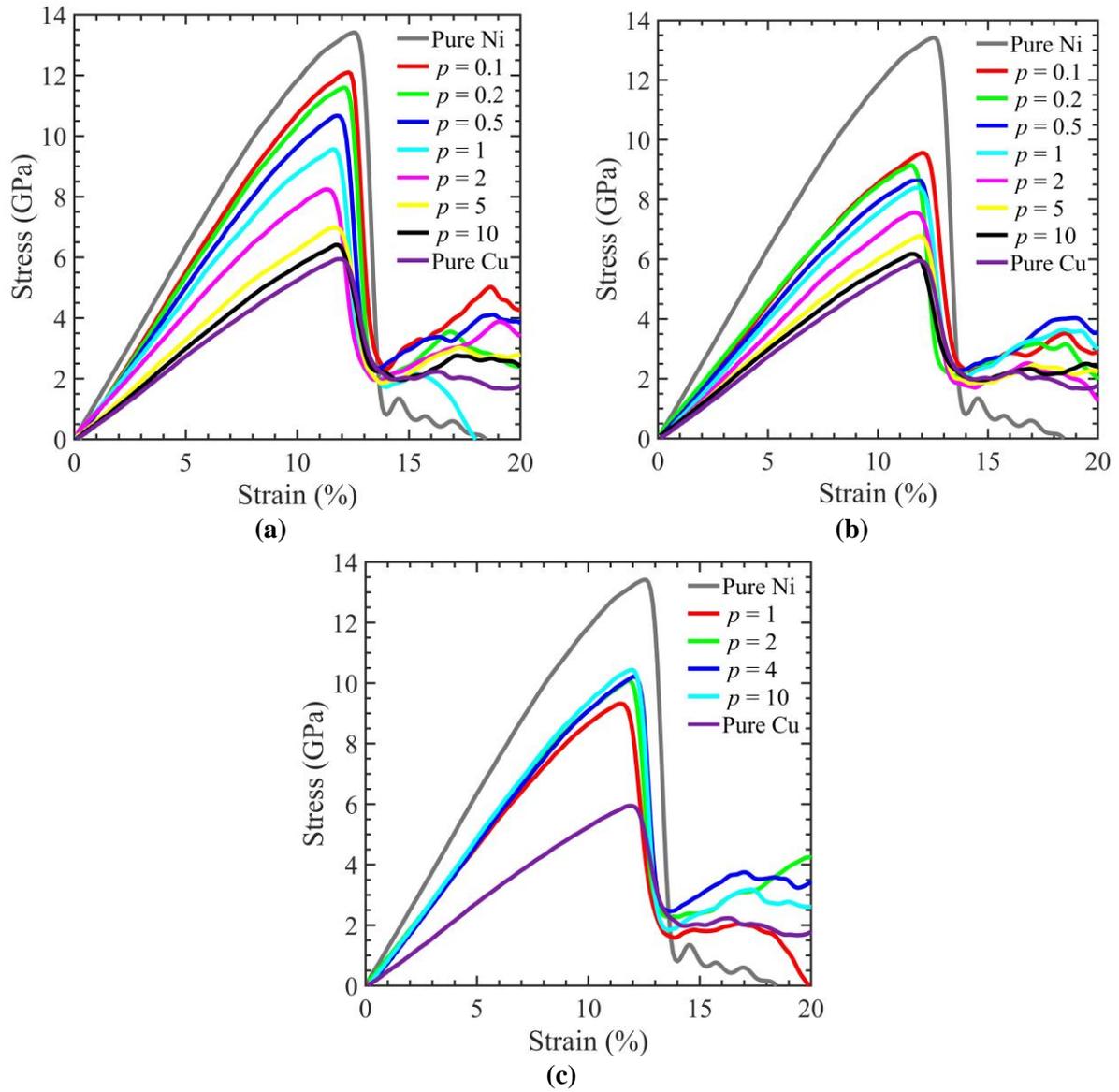

**Figure 4.** Stress-strain curves of **(a)** P-FGM, **(b)** E-FGM and **(c)** S-FGM NWs for different values of function parameter *p*.

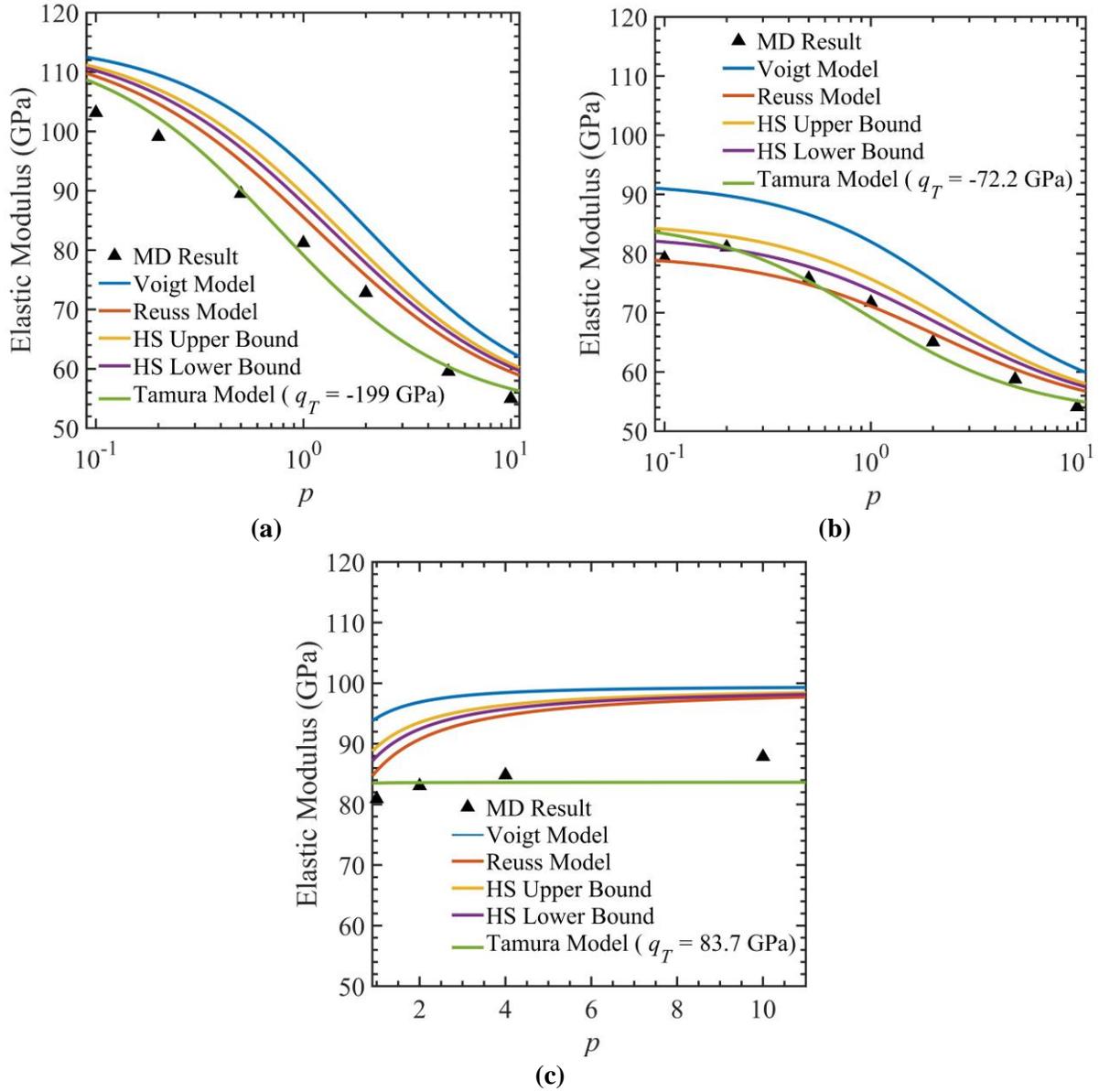

**Figure 5.** Variation of Elastic moduli of **(a)** P-FGM, **(b)** E-FGM and **(c)** S-FGM NWs for different values of function parameter, $p$, and comparisons with the Voigt model, Reuss Model, Hashin–Shtrikman (HS) upper and lower bounds and the Tamura model with appropriate values of $q_T$.

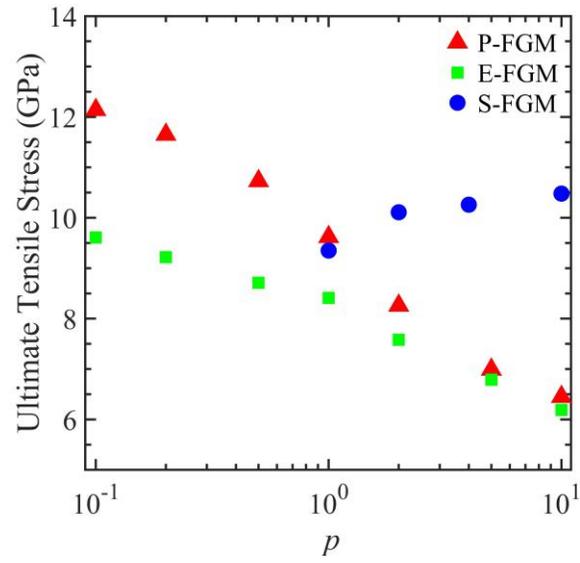

**Figure 6.** Variation of Ultimate tensile stress of P-FGM, E-FGM and S-FGM NWs for different values of function parameter, $p$.

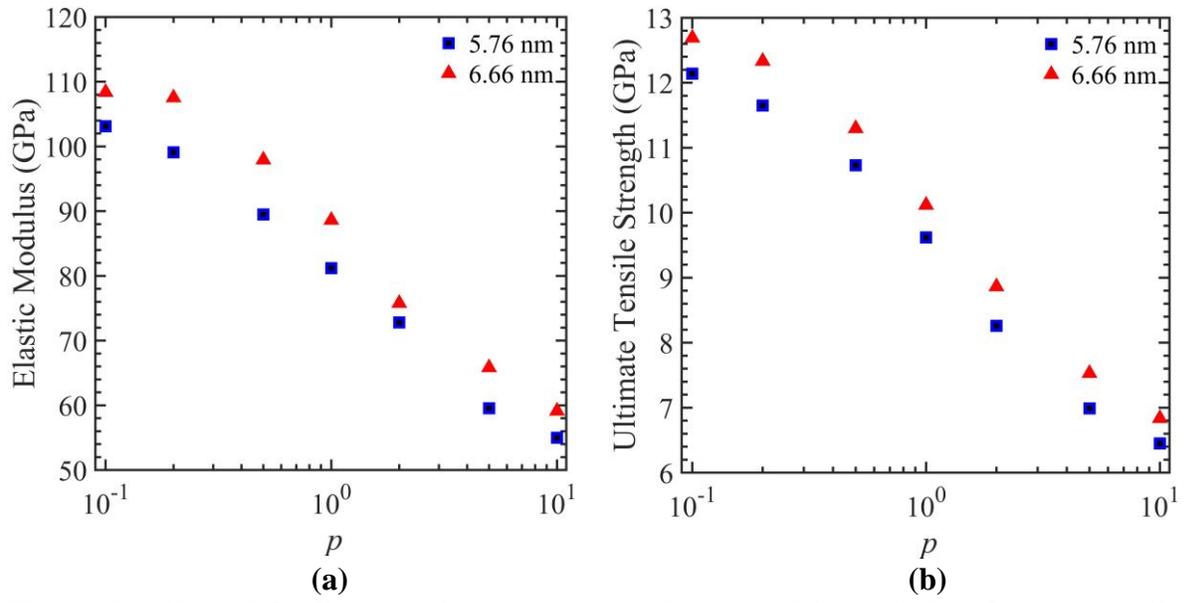

**Figure 7.** Effect of P-FGM NW diameter on **(a)** Elastic Modulus and **(b)** Ultimate Tensile Strength, at different values of function parameter, *p*.

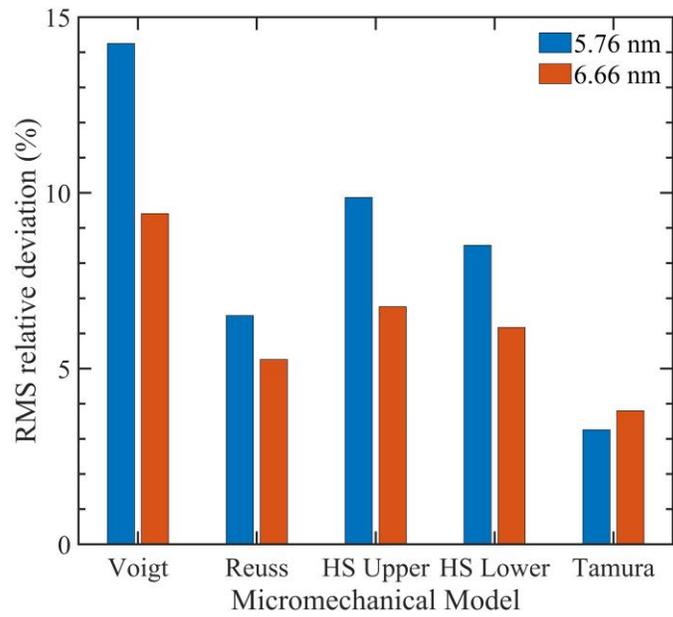
**Figure 8.** Effect of P-FGM NW diameter on the RMS relative deviations of MD results from different micromechanical models.

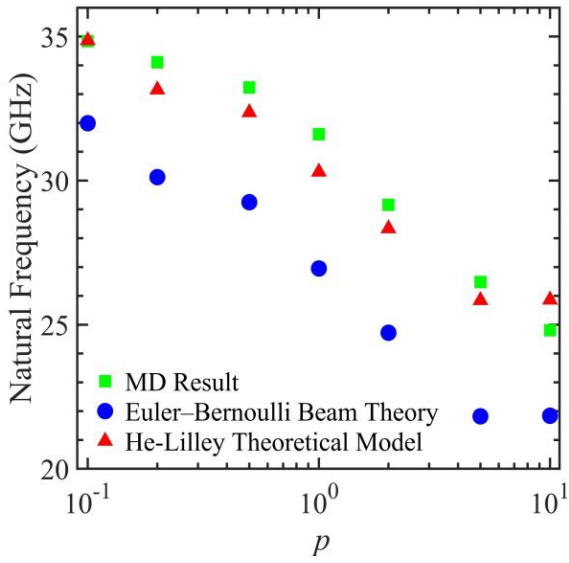
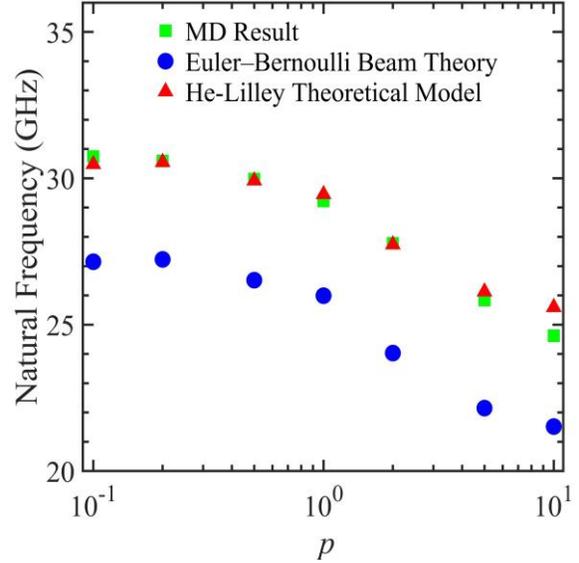

(a)                                    (b)

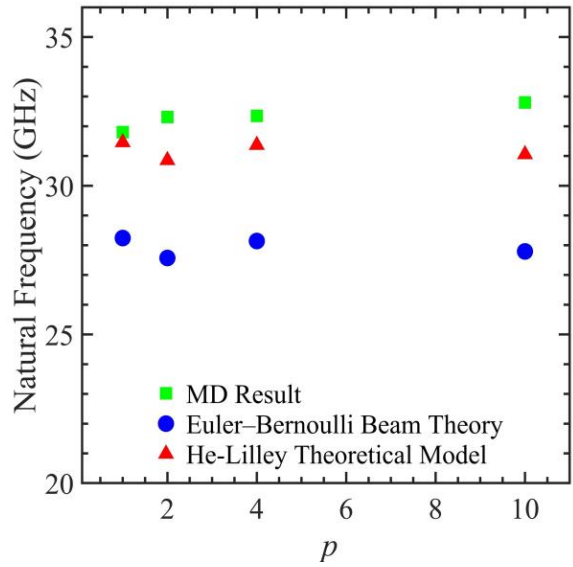

(c)

**Figure 9.** Variation of Natural Frequency of (a) P-FGM, (b) E-FGM and (c) S-FGM NWs for different values of function parameter, *p*, and comparison with the predictions of Euler-Barnoulli Beam Theory and He-Lilley theoretical model.

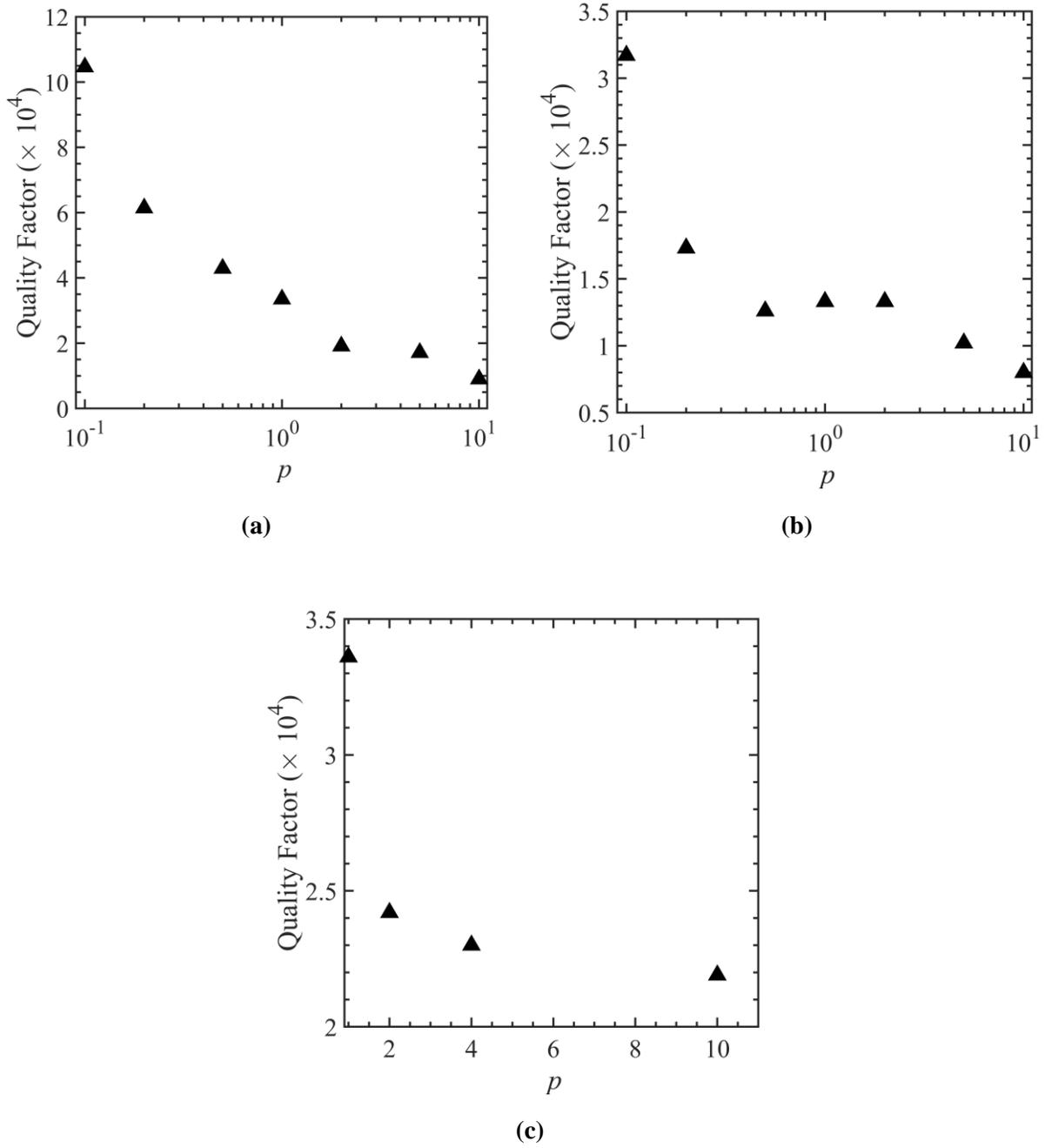

**Figure 10.** Variation of Q-factor of (a) P-FGM, (b) E-FGM and (c) S-FGM NWs for different values of function parameter, $p$.

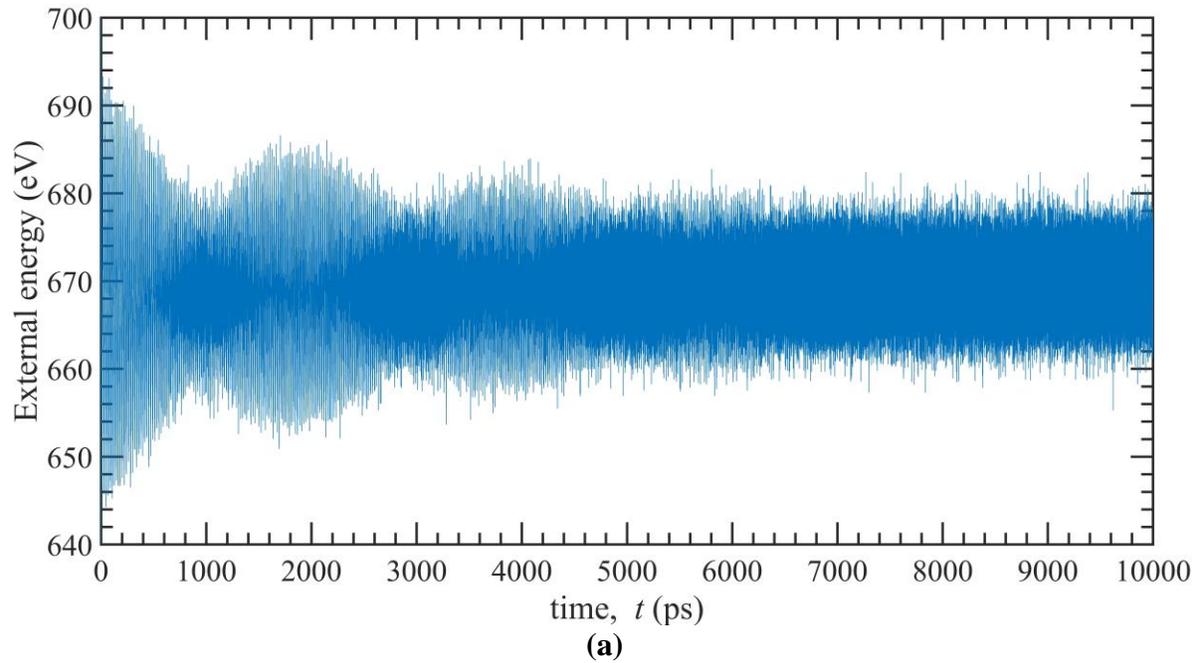

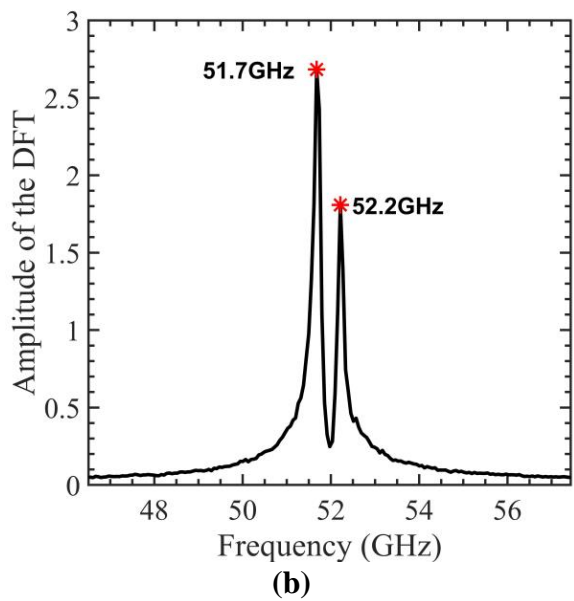
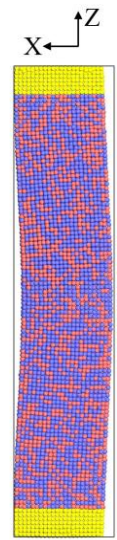
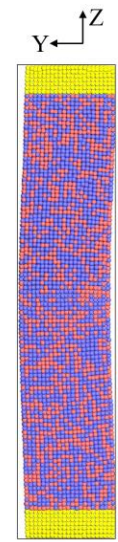

**Figure 11. (a)** Time history *EE* during vibration, **(b)** one-sided FFT power spectrum of *EE* in frequency domain, atomic configuration at 900 ps to visualize vibration along **(c)** *x*-direction and **(d)** *y*-direction, for E-FGM, $p = 5$ NW.